# An improved method to measure microwave induced impulsive forces with a torsion balance or weighing scale

Chris P. Duif
Faculty of Applied Sciences, Delft University of Technology,
and
Reactor Institute Delft, Delft University of Technology,
Mekelweg 15, 2629 JB, Delft, The Netherlands

e-mail: c.p.duif@tudelft.nl   or   chrisd@space-time.info

June 15, 2017

**Abstract**
A novel method is presented for measuring impulsive forces generated by devices which are fed with medium power microwave signals. The forces are measured with a torsion balance or weighing scale, as usual, but the microwave signal is coupled directly to the device under test via a special coupling cavity instead of being generated on the measurement device. The method was verified at power levels up to 15 W, has a transmission of ≥75% (−1.3 dB attenuation) and is shown not to exert disturbing forces at this power level (< $10^{-5}$ N).
The application of this way of supplying microwave signals could significantly improve experiments which otherwise suffer from heat dissipation and Lorentz forces by components present on the force measurement device. One of the applications might be to answer the question whether the so-called EmDrive works or not, which question, despite serious work in the past few years, has not been satisfactory answered yet.

**Introduction**

Several reports have been published in which anomalous forces are reported when conical or tapered microwave cavities are excited at their resonance frequencies (usually in the range 2–3 GHz). These devices were called *EMDrives* by the inventor who first published about it, Roger Shawyer [1, 2]. Shawyer claims that they can be used for space propulsion and have therefore attracted quite some attention [3, 4].
However, these devices, as being closed systems and still producing thrust, would clearly violate conservation of momentum and Newton's Third Law and are therefore met with a lot of scepticism [5].
Several researchers have tried to replicate Shawyer's experiments, with mixed results. Some of them had an undecided outcome [6], but in 2016 a report confirming the existence of anomalous forces was published by White *et al.* [7] of the so-called *Eagleworks Laboratories*, sponsored by NASA. In this report several shortcomings of previously published experiments were avoided. One of the improvements was that is was conducted in a vacuum, in that way avoiding disturbing effects of air currents. Still, this study has attracted substantial critiques, *e.g.* by Hathaway [8].

So, after more than a decade of research in this field it is still very controversial and far from being proved or disproved. The main sources of uncertainties in these experiments come from the relative big amount of heat generated on the measurement device itself (microwave amplifier and other components, dissipated heat by the cavity), the Lorentz forces of DC currents [9] and (in some experiments) the attached power cables (Shawyer) or liquid metal contacts (Eagleworks Lab [7, 10], Tajmar & Fiedler [6]). Also, the use of air tracks or air bearings in some of these measurements (Shawyer) is highly disputed [11].

In this study, a method is described, in which the microwave signals are generated outside the torsion balance (or weighing scale) setup and supplied through a special coupling cavity. The cavity essentially is a tuneable microwave bandwidth filter of which one of the endplates is not mechanically in contact with the cavity. This enables a contactless supply of microwave power.

The method was tested for a possible disturbing influence, and it was shown that the attraction between the two parts is very small and therefore is not expected to influence torsion balance or weighing scale measurements.

The advantages of this method are: the heat generation on the measurement device will be one or two orders of magnitude smaller; no DC currents (Lorentz forces), except for the microwave power detector, no need for magnetic materials (*e.g.*, a microwave circulator) on the balance; a substantial reduction of the mass, by several kg's, which improves the response time of the measurement device; easier control of the frequency and power (no need for wireless control); no need for special amplifiers when measuring in a vacuum.

**Setup**

As stated above, the coupling cavity basically is a cylindrical microwave cavity bandwidth filter, with B-loops, as exiting and pick-up coil, respectively, on both endplates. One of the endplates is positioned contactless inside the cavity. In this way, it enables practically forceless power transfer to the measurement setup (*e.g.*, torsion balance, weighing scale) and also enables to tune the pass-band frequency to the desired value (Fig. 1).

It was chosen to test pass-bands in the range 3–4 GHz, which led to using a cavity with an inner diameter of 98 mm. The (copper) wall thickness is 0.6 mm.

B-loops were developed which give a relative broad pass-band. They are made of semi-rigid cable (RG402) with a loop diameter of 15.0(5) mm and have a separation from the endplate of 32.05(5) mm, see Fig. 2. The wire thickness is that of the RG402 inner conductor, 0.9 mm.

These coils were connected with SMA-cables to the microwave amplifier and the device-under-test (or a detector), respectively (see Fig. 1).

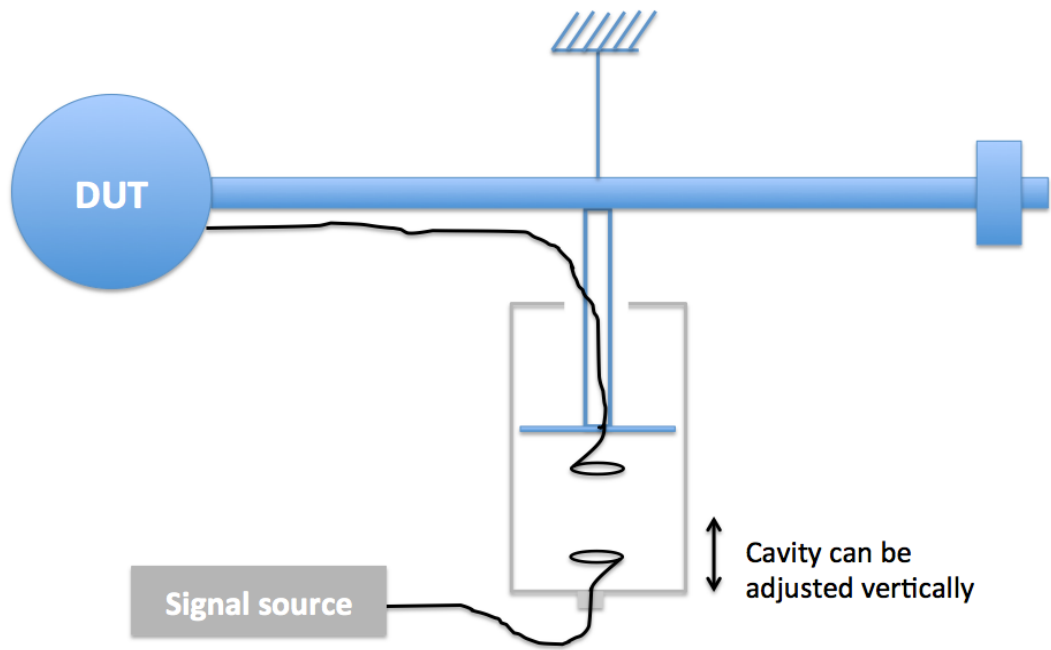

**Fig. 1.** Schematic drawing of the coupling cavity mounted under the centre of a torsion balance. Exciting and pick-up loops and coaxial cables in black, torsion balance and parts attached to this in blue, parts fixed "to the floor" (the cavity ) in grey. Of course, in principle, the cavity can also be attached to the pendulum, but the cavity walls are the parts which heat up the most due to the wall currents. DUT = the *device under test*.

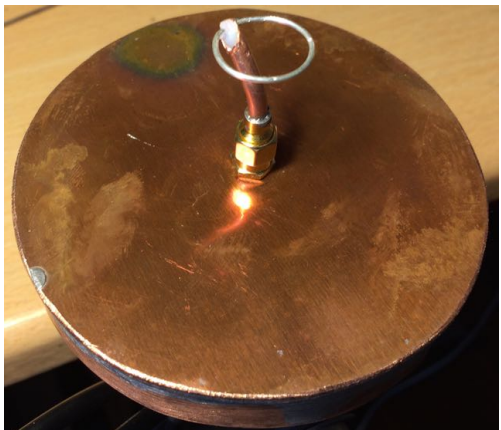

**Fig. 2.** The B-loop on the movable endplate.

The diameter of the movable endplate is 94 mm, leaving a gap of 2 mm between this endplate and the cavity wall. It is mounted on a rod which is used to position it inside the cavity. The cavity length in this way can be varied from ~60–160 mm (the total length of the copper tube is 220 mm). An extra end cap with a hole of 30 mm closes the space behind the movable endplate (see Fig. 3). It is through the gap between this hole and the supporting rod that microwave power will leak out.

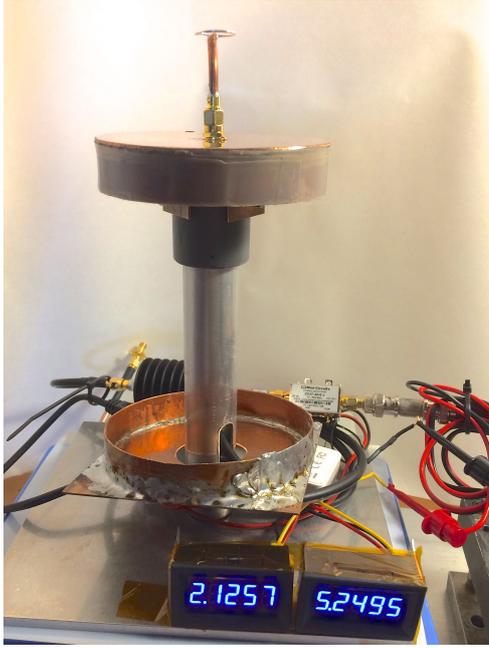

**Fig. 3.** The movable endplate with pick-up coil on a platform with power meter and readout. The display on the right monitors the battery voltage.

The resonance frequency dependence on the cavity length $l_c$ can be calculated from:

for TM modes

$$f_{mnp} = \frac{c}{2\pi\sqrt{\mu_r \, \epsilon_r}} \sqrt{\left(\frac{X_{mn}}{R}\right)^2 + \left(\frac{p\pi}{l_c}\right)^2}$$

for TE modes

$$f_{mnp} = \frac{c}{2\pi\sqrt{\mu_r \, \epsilon_r}} \sqrt{\left(\frac{X'_{mn}}{R}\right)^2 + \left(\frac{p\pi}{l_c}\right)^2}$$

in which $X_{mn}$ is the *n*-th zero of the *m*-th Bessel function, and $X'_{mn}$ is the *n*-th zero of the *derivative* of the *m*-th Bessel function, R is the cavity radius and $l_c$ is the cavity length.
Therefore, in order to be tunable, a mode with p≥1 has to be chosen.

In the range 2800–4000 MHz, 3 usable resonances with a broad pass-band have been identified for the tested cavity, they are listed in Table 1 for a cavity length $l_c$ of 98.0(5) mm). These frequencies depart from the calculated values with the equations above, due to the coupling loops (the large distance from the endplate, among others) and possibly the interaction with 'the other cavity' (between moving endplate and end cap).

**Microwave power transfer of the coupling cavity**

For most of the applications, it is desirable that the pass-bands have a width of at least a few MHz. In the EmDrive experiments, for example, a cavity has to be exited, which resonance frequency will change in time due to temperature changes [7, 10]. This turned out to be easy to achieve with the right coupling loops. As can be seen in the network analyser measurements below, the developed coupling cavity has bandwidths of some tens of MHz (with approx. 1 dB attenuation). These pass-bands can be shifted by changing the cavity length (the '3600 MHz resonance' for example has a range of approx. 3350–3700 MHz).

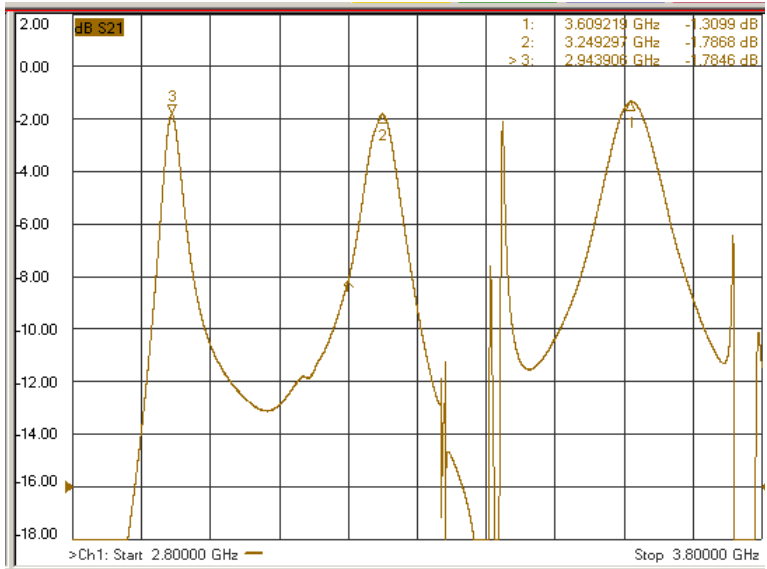

**Fig. 4.** Transmission ($S_{21}$) of the coupling cavity, $l_c$ = 98.0(5) mm, 2800–3800 MHz. Relative broad pass-bands at 3609, 3249 and 2944 MHz.

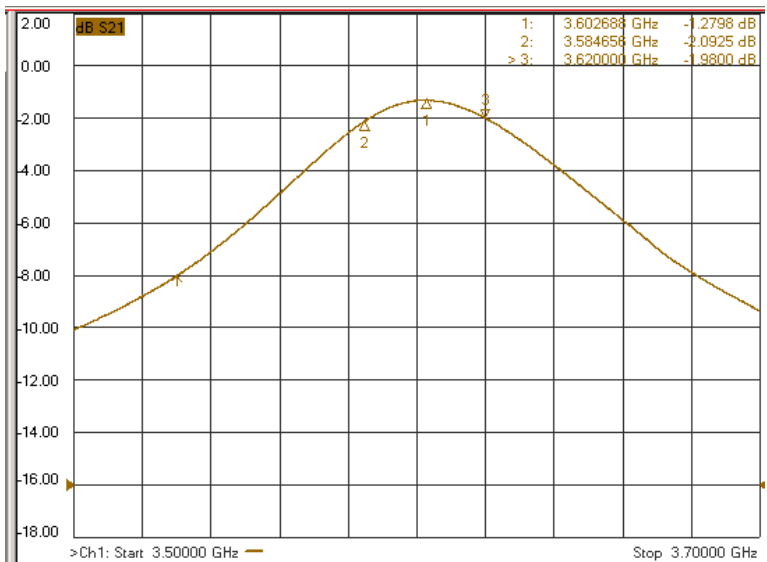

**Fig. 5.** The '3600-MHz resonance' in more detail, $l_c$ = 98.0(5) mm. The −1 dB width from the top (1.28 dB) is ~60 MHz.

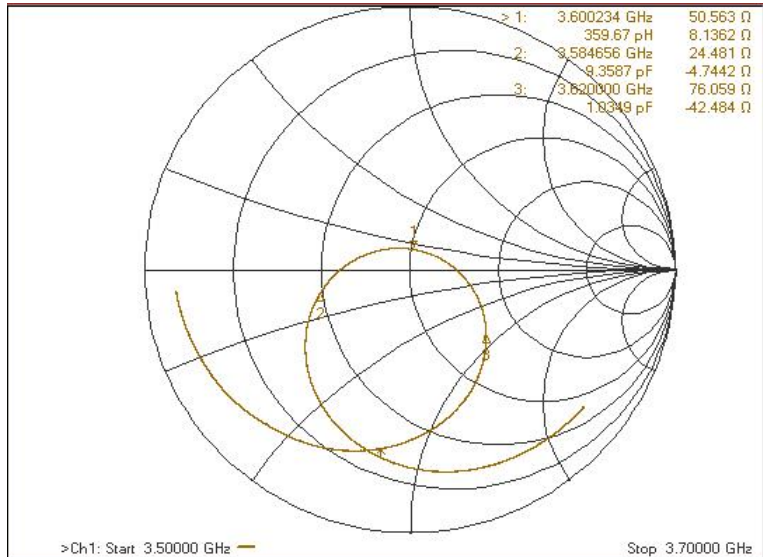

**Fig. 6.** Smith chart of an $S_{11}$ measurement of the coupling cavity, with $l_c$ = 98 mm, of the '3600-MHz resonance'. Quite near 50Ω, as can be seen.

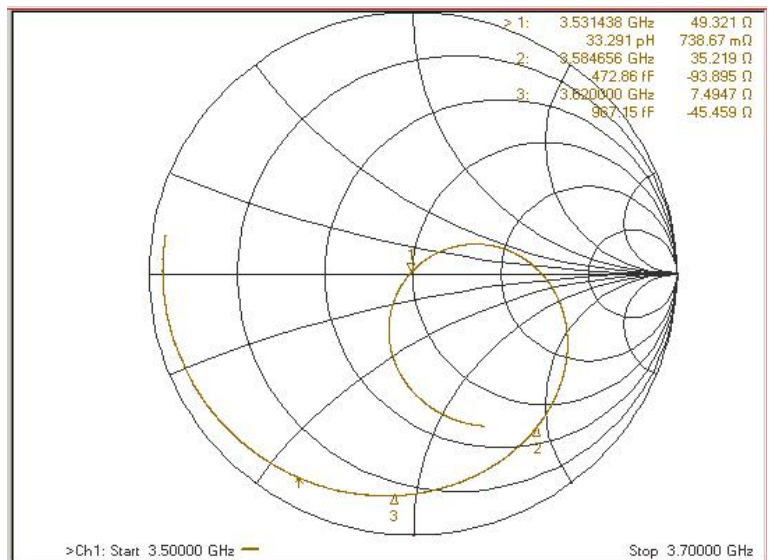

**Fig. 7.** Smith chart of an $S_{11}$ measurement of the coupling cavity, with $l_c$ = 106 mm, of the '3600-MHz resonance'. Almost exactly 50Ω (cursor 1).

**Table 1.** The pass-bands in the range 2800–4000 MHz.

| Frequency @ $l_c$=98 mm [MHz] | Mode | Range [MHz] | Minimum damping [dB] |
|---|---|---|---|
| 3617 | TE112? | 3350–3700 | 1.3 |
| 3251 | TE211? | 3150–3260 | 1.5 |
| 2948 | TM011? | 2750–2970 | 1.7 |

**Checking for forces between the two cavity parts**

A check for possible disturbing vertical forces was performed with a sensitive scale (Mettler PM1200, resolution 1 mg, maximum load 1.2 kg). The movable endplate and supporting column of the cavity was mounted on a plate on which also the battery-fed RF power meter (Mini-Circuits ZX47-40+) and the read-out display was mounted. This setup is depicted schematically in Fig. 8, its electronic scheme in Fig. 9, the actual realisation can be seen in Fig. 10.

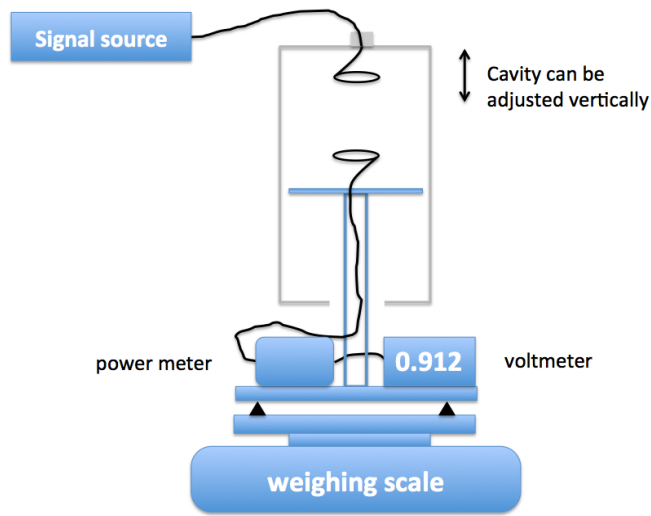

**Fig. 8.** Weighing of the movable part of the coupling cavity, schematically, in order to check for (vertical) disturbing forces during power transfer. Details like RF attenuators and battery are left out.

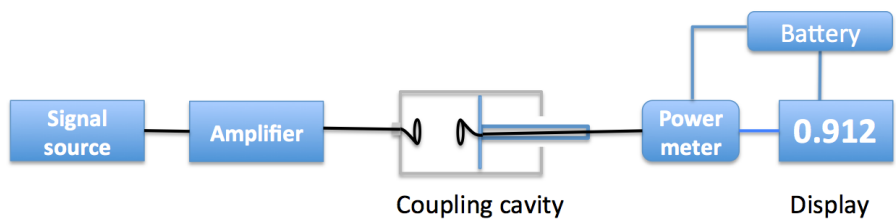

**Fig. 9.** Electronic scheme of the power transfer and measurement when measuring on the weighing scale. The microwave power detector was a Mini-Circuits ZX47-40+ with 45 dB attenuator.

Vertical forces can have a disturbing influence on a torsion balance, *e.g.*, if the coupling pipe (movable endplate) is not located exactly under the centre of the balance (torsion wire).

It was measured that at the maximum applied power, 15 W (~42 dBm), and at 3650 MHz, the weight difference was less than 1 mg  =>  $F_{vert} < 10^{-5}$ N.

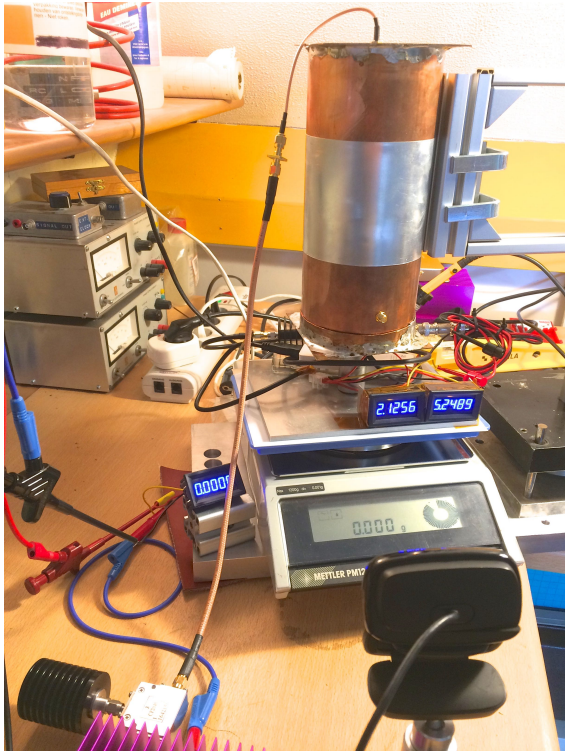

**Fig. 10.** Measurement of vertical forces with the movable endplate of the cavity mounted on a platform. On this platform, also the power meter, battery and voltmeters with display are mounted. In the lower left corner, a part of the microwave amplifier can be seen. The display on the left side of the scale shows the reflected power.

Preliminary measurements with a torsion balance setup show that disturbing torque on the balance is very small (< 1 µNm), but this will be researched in more detail in the near future.

**Using the coupling cavity on a torsion balance**

The coupling cavity described in this article will be used to drive a conical shaped microwave cavity (*e.g.*, for the TE013-mode at 3.6 GHz => big/small diameter = 203/121 mm, length = 163 mm).

For these experiments, it is necessary to monitor forward as well as reflected power and to read these out in a contactless way. In Fig. 11 a setup is schematically depicted in which a directional coupler and two battery-fed power meters are used.

If the dissipated heat by the wall of the coupling cavity is a problem, the excess heat can easily be removed with a water-cooled mantle.

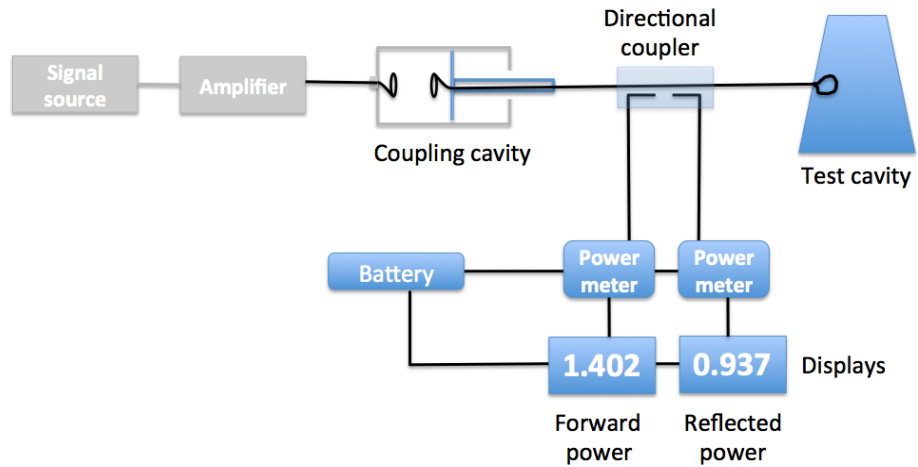

**Fig. 11.** Schematic of the coupling cavity applied to a torsion balance setup for an EmDrive experiment (parts on the balance in blue, parts 'on the ground' in grey). The power delivered to the conical cavity, and the reflected power, are monitored with a directional coupler and two power sensors with optical readout (displays monitored with a webcam). Some details, like RF attenuators and circulator, are left out.

**Microwave leakage, shielding**

The gap around the movable endplate and the hole in the 'end cap' (around the supporting rod on which this endplate is mounted), make that a considerable amount of microwave power radiates outward. It is estimated that this can be a few watt at a supplied power of 20 W. The space between the movable endplate and the end cap of the copper pipe form another cavity resonator which is coupled to the coupling cavity. Outward radiation of the whole device can be reduced by a clever choice of the dimensions of this extra cavity and the use of microwave absorbing materials (*e.g.*, on the walls of the space behind the movable endplate). The coupling cavity presented in this report is far from optimal in this respect. The radiated microwave waves could in principle disturb the electronics of the balance, but that has not been observed in this setup.

**Conclusions**

It is possible to transfer microwave power contactless to a force measuring device with an efficiency of >75% (less than 1.3 dB loss). By further optimising of the coupling loops, this probably can even be improved by something of the order of 0.5 dB. The practical limits for making such a coupling cavity will probably lie in the range of a few 100 MHz to 20 GHz and powers of up to a few hundred watt. Experiments in a particular pass-band of the used cavity show that no noticeable disturbing (electromagnetic) forces are exerted on the force measuring device. Further development to reduce the radiated microwave power will be very useful.

**Acknowledgements**

I wish to thank Fred van der Zwan of the Faculty of EWI, TU Delft, for assisting in the network analyser measurements and kind hospitality.